\documentclass[pre,a4paper,onecolumn,notitlepage,nofootinbib]{revtex4-1}
\usepackage[cp1250]{inputenc}
\usepackage[T1]{fontenc}
\usepackage[dvips]{graphicx}
\usepackage{amsmath}
\usepackage{amsfonts}

\newcommand{\uni}[1]{\,\mathrm{#1}}
\newcommand{\beq}{\begin{equation}}

\newcommand{\eeq}{\end{equation}}

\begin{document}
\title{\bf Maximal entropy random walk in community finding}

\author{J.K. Ochab}\email{jeremi.ochab@uj.edu.pl}

\author{Z. Burda}\email{zdzislaw.burda@uj.edu.pl}

\affiliation{Marian Smoluchowski Institute of Physics and 
Mark Kac Complex Systems Research Center \\
Jagiellonian University, Reymonta 4, 30-059 Krak\'ow, Poland}

\begin{abstract}
The aim of this paper is to check feasibility of using the maximal-entropy random walk in algorithms finding communities in complex networks.
A number of such algorithms exploit an ordinary or a biased random walk for this purpose.
Their key part is a (dis)similarity matrix, according to which nodes are grouped.
This study encompasses the use of the stochastic matrix of a random walk, its mean first-passage time matrix, and a matrix of weighted paths count.
We briefly indicate the connection between those quantities and propose substituting the maximal-entropy random walk for the previously chosen models.
This unique random walk maximises the entropy of ensembles of paths of given length and endpoints, which results in equiprobability of those paths.
We compare performance of the selected algorithms on LFR benchmark graphs.
The results show that the change in performance depends very strongly on the particular algorithm, and can lead to slight improvements as well as significant deterioration.

\medskip


\end{abstract}

	\maketitle
		
\section{Introduction}

Relationships between entities can be represented as a graph structure upon which some process takes place,
be it information or opinion spread on social networks, including citation and collaboration networks, WWW or the Internet,
or perhaps a physical process (molecule movement) on physical or biological networks.
One of the natural question to ask is whether there are groups of entities which are connected stronger to each other than to the rest of the network.
Due to the sociological legacy, these are called \emph{communities},
but can comprise of researchers, websites, genes or transcription factors alike.

A plenitude of methods have been devised to find such communities,
and a plenitude of definitions have been conceived to tell what it is that we really look for.
These definitions and methods have been thoroughly reviewed in \cite{Fortunato}.
A particular subgroup of the algorithms is based on random walks (RWs),
since intuitively a random walker is expected to spend a longer time inside well-connected graph regions,
and there should be only a slim chance that it crosses from one to another.

The most common choice for such algorithms has been the well-known random walk
defined by equal probabilities of going from a node to any of its nearest neighbours,
which we call the generic random walk (GRW).
On the contrary, maximal-entropy random walk (MERW) ensures equiprobability of all paths of a given length and endpoints.
Although for many problems GRW and biased RWs are often more suitable, MERW deserves particular interest:
while the former maximises entropy locally (entropy of the nearest neighbour selection),
the latter maximises entropy globally (entropy of the path selection) \cite{ZB1,ZB2}.
Among its curious behaviours, MERW exhibits localization of its stationary distribution on diluted lattices \cite{ZB1,ZB2,BW} and Cayley trees \cite{ZB3,Demo1}, it also relaxes extremely fast on these trees \cite{ZB3,Demo2}, while it does very slowly between two identical connected regions \cite{JO}.
Thus, we believe MERW can serve alongside GRW as a null model of random processes on networks.

It is noteworthy that equiprobable paths (as generated by MERW) are the natural choice for an ensemble used in Feynman path integrals (e.g. discrete quantum gravity models with curved space-time) \cite{ZB2} or in the optimal sampling algorithm in the path-integral Monte Carlo methods \cite{H}. Entropy maximization is a global principle much like the least action principle. It has earlier led to the biological concept of evolutionary entropy \cite{Evolutionary1}. Interestingly, the value of entropy for a given graph, as defined by MERW, has been found useful for selection of robust networks \cite{Evolutionary2}. 
Finally, it has begun to be used in the study of complex networks \cite{MERW+CN1,Delvenne,MERW+CN3,MERW+CN4,MERW+CN5}.

\section{Generic and maximal-entropy random walks}
Let us consider a discrete time random walk on a finite connected undirected graph,
with its stochastic matrix $\mathbf{P}$ is constant in time. 
An element $P_{ij} \ge 0$ of this matrix encodes the probability that a walker that stands on a node $i$ 
at time $t$ hops to a node $j$ at time $t+1$. These matrix elements
fulfil the condition $\sum_j P_{ij} = 1$ for all $i$, which means
that the number of walkers is conserved.
An additional assumption allows the walkers to hop only to a neighbouring node.
This can be formulated as $P_{ij}\leq A_{ij}$,
where $A_{ij}$ is the corresponding element of the adjacency matrix $\mathbf{A}$ of the graph: 
$A_{ij}=1$ if $i$ and $j$ are neighbours, and $A_{ij}=0$ otherwise.

For any time $t$, the probability of a walker staying on a given vertex of the graph
is encoded in the vector $\mathbf{\vec{\pi}}(t)=(\pi_1(t),\ldots,\pi_N(t))^T$.
The initial distribution of particles is $\mathbf{\vec{\pi}}(0)$,
and the distribution after $t$ steps $\mathbf{\vec{\pi}}(t)^T=\mathbf{\vec{\pi}}(0)^T\mathbf{P}^t$.
A quantity of interest is the stationary probability distribution, which we assume exists.
Then it is given by a solution of
\beq
\mathbf{\vec{\pi}}^T=\mathbf{\vec{\pi}}^T\mathbf{P},
\eeq
and may be regarded as the probability distribution after infinite time.

GRW is realised by the following stochastic matrix:
\beq
P_{ij} = \frac{A_{ij}}{k_i} \ ,
\label{Porw}
\eeq
where $k_i = \sum_j A_{ij}$ denotes the node degree.
The factor $1/k_i$ in the above formula
produces uniform probability of selecting one of $k_i$ neighbours of the node $i$.
This choice maximises the entropy of neighbour selection
and corresponds to the standard Einstein-Smoluchowski-Polya random walk. 
The stationary probability distribution of GRW is given by $\pi_i = k_i/\sum_j k_j$.

The other type of random walk, MERW,
is defined by a stochastic matrix that maximises entropy of a set of trajectories with a given length and end-points.
This is a global principle similar to the least action principle.
It leads to the following stochastic matrix:
\beq
\label{Pmerw}
P_{ij} = \frac{A_{ij}}{\lambda_0} \frac{\psi_{0j}}{\psi_{0i}} \ ,
\eeq
where $\lambda_0$ is the largest eigenvalue of the adjacency matrix $\mathbf{A}$,
and $\psi_{0i}$ is the $i$-th element of the corresponding eigenvector $\mathbf{\vec{\psi}}_0$.
By virtue of the Frobenius-Perron theorem all elements of this vector are of the same sign,
because the adjacency matrix $\mathbf{A}$ is irreducible.
For a stochastic matrix to maximise the entropy of an ensemble of paths the choice \eqref{Pmerw} is unique.

The defining condition of entropy maximization leads to equiprobability of paths. More precisely, let us take a sequence of nodes $\gamma_{a_0 a_\tau}=(a_0,a_1,\ldots,a_\tau)$, which is a path of $\tau$ steps with the initial node $a_0$ and the final node $a_\tau$. The probability of visiting this sequence of nodes is
\beq
P(\gamma_{a_0 a_\tau})=P_{a_0 a_1}P_{a_1 a_2}\cdots P_{a_{\tau-1} a_\tau} \ ,
\eeq
which results from the Markov property of the random walk.
Upon substitution of MERW's stochastic matrix one obtains
\beq
P(\gamma_{a_0 a_\tau})=\frac{1}{\lambda_0^\tau}\frac{\psi_{0a_0}}{\psi_{0a_\tau}} \ ,
\eeq
which depends only on the number of steps and the two ending points, but is independent of the intermediate nodes. This is what we mean by equal probability of paths of a given length and end-points. Consequently, the probability measure on this ensemble of paths is uniform, and its entropy is maximal.

The stationary state of MERW is given by Shannon-Parry measure \cite{P}:
\beq
\label{statP}
\pi_i = \psi^2_{0i} \ .
\eeq
The last formula forms a connection between MERW and quantum mechanics,
since $\psi_{0i}$ can be understood as the wave function of the ground state of 
the operator $-\mathbf{A}$ and $\psi^2_{0i}$ as the probability 
of finding a particle in this state \cite{ZB1,ZB2}.
The two types of random walk, (\ref{Porw}) and (\ref{Pmerw}), are behave identically on $k$-regular graphs.
In general, however, they have completely disparate properties.


\section{(Dis)similarity matrices for community finding algorithms}
\label{sec:theory}

Methods of both assessing centrality \cite{White} and finding communities \cite{Harel,Latapy} have widely utilised calculating powers of the stochastic matrix.
The one by Latapy and Pons \cite{Latapy} uses the dissimilarity matrix
\beq
\label{rdist}
r(t)_{ij}=\sqrt \frac{\sum_{k}[(\mathbf{P}^t)_{ik}-(\mathbf{P}^t)_{jk}]^2}{\pi_k},
\eeq
where the division by $\pi_k$ is supposed to reduce the effect of a vertex's centrality.
Originally, $\mathbf{P}$ and $\mathbf{\vec{\pi}}$ corresponding to GRW were chosen.

Another approach is an explicit use of the mean first-passage times (MFPT)\cite{Zhou1,Zhou2,Zhou3}.
MFPT matrix $\mathbf{M}$ is a useful and well-studied quantity characterising RWs. 
Its construction with the use of the fundamental matrix $\mathbf{Z}$ is given in \cite{Snell1,Snell2}
\begin{eqnarray}
\label{eq:MFPT}
\mathbf{Z}&=&(\mathbf{1}-\mathbf{P}+\mathbf{\vec{e}}\mathbf{\vec{\pi}}^T)^{-1}\\
\mathbf{M}&=&(\mathbf{E}\mathbf{Z}_{d}-\mathbf{Z})\mathbf{D} \quad,
\end{eqnarray}
where $\mathbf{1}$ is the identity matrix, $\mathbf{\vec{e}}=(1,1,...,1)^T$, $\mathbf{E}$ is a matrix of all ones, $\mathbf{Z}_{d}$ is a diagonal matrix with elements $(\mathbf{Z}_{d})_{ii}=Z_{ii}$, and $\mathbf{D}$ is a diagonal matrix with elements $(\mathbf{D})_{ii}=1/\pi_{i}$.
The elements $M_{ij}$ encode the average time to reach the vertex $j$ from $i$ for the first time (in general $M_{ij}\neq M_{ji}$).

The last approach we discuss is a similarity matrix containing the average number of paths between two given nodes (which is just $\mathbf{A}^t$)
with weights that depend on the paths' length
\beq
\label{Gmu}
\mathbf{G}(\mu)=\sum_{t=0}^{\infty} e^{-\mu t} \mathbf{A}^t.
\eeq
For $e^{\mu}\equiv \lambda>\lambda_0$ the sum is convergent and can be carried out with the use of spectral decomposition of $\mathbf{A}$.
From the point of view of paths' statistics, $\mathbf{G}(\mu)$ defines the grand-canonical ensemble of paths.
An element $G_{fi}(\mu)$ corresponds to the grand canonical partition function, $\mu$ to the chemical potential,
and the average path length is $\langle t \rangle_{fi} = -(\ln G)'_{fi}(\mu)$. To avoid conflicting notation, henceforth we use $\lambda \equiv e^{\mu}$, whereas the symbol $\mu$ will be exclusively reserved for the mixing parameter of benchmark graphs (see Sec.\ref{sec:bench}).

In the case of MERW and GRW (generally, for any RW for which  $\mathbf{D}^{-1/2}\mathbf{P}\mathbf{D}^{1/2}$ is symmetric)
it can be shown that these three quantities are intimately related constituting a common framework for a number of centrality measures \cite{ZB2,JO1}.


\section{Comparison of community finding algorithms}
\label{sec:community}

Each of the above quantities has an analogic centrality measure: $\mathbf{r}$ has the stationary state centrality and centralities defined by summation of powers of the stochastic matrix, $\mathbf{G}$ has the eigenvector centrality and centralities defined by path enumeration, and $\mathbf{M}$ has a centrality defined by the inverse of its average rows \cite{JO1}. These are natural counterparts to some community finding methods.

Just as centrality may be defined with the use of the principal eigenvector of the adjacency matrix or the stochastic matrix
(then the eigenvector is the stationary state),
there is a family of community finding methods analysing the rest of the eigenvectors
(often it is the spectrum of Laplacian that is analysed) \cite{Donetti,Simonsen1,Simonsen2,Shi,Shi1,Caldarelli}. 
However, having the two random walks at hand, we are more interested in methods that utilise their characteristics.
Particularly, we try to assess what difference it makes, when we switch between those two random walks.

There are a number of methods using powers of the transition matrix.
For instance, \cite{Harel} use the matrix
\beq
\label{eq:harel}
\mathbf{P}^{\leq T} \equiv\sum_{t=1}^{T}\mathbf{P}^t,
\eeq
where $\mathbf{P}$ corresponded to GRW, and $T$ was taken around $2-3$.
The assumption is that two nodes are close to each other if the corresponding rows of $\mathbf{P}^{\leq t}$ matrix are similar.
One of the proposed similarity functions between two vectors is
\beq
\label{eq:harel1}
\uni{sim}(\mathbf{\vec{x}},\mathbf{\vec{y}})=\exp\left(2T-\sum_{i=1}^{N}|x_i-y_i|\right)-1.
\eeq
In this formula, if $T=1$ the vectors $\mathbf{\vec{x}},\mathbf{\vec{y}}$ are rows of the stochastic matrix,
hence the elements of each of them sum up to $1$.
There are $T$ stochastic matrices summed in \eqref{eq:harel}, hence in general the elements of each vector sum up to $T$.
If the two vectors are maximally different, the sum in \eqref{eq:harel1} becomes $2T$, and the similarity reaches the lower boundary value of $0$.

\begin{figure}[hp]
	\centering
		\includegraphics[width=0.47\textwidth]{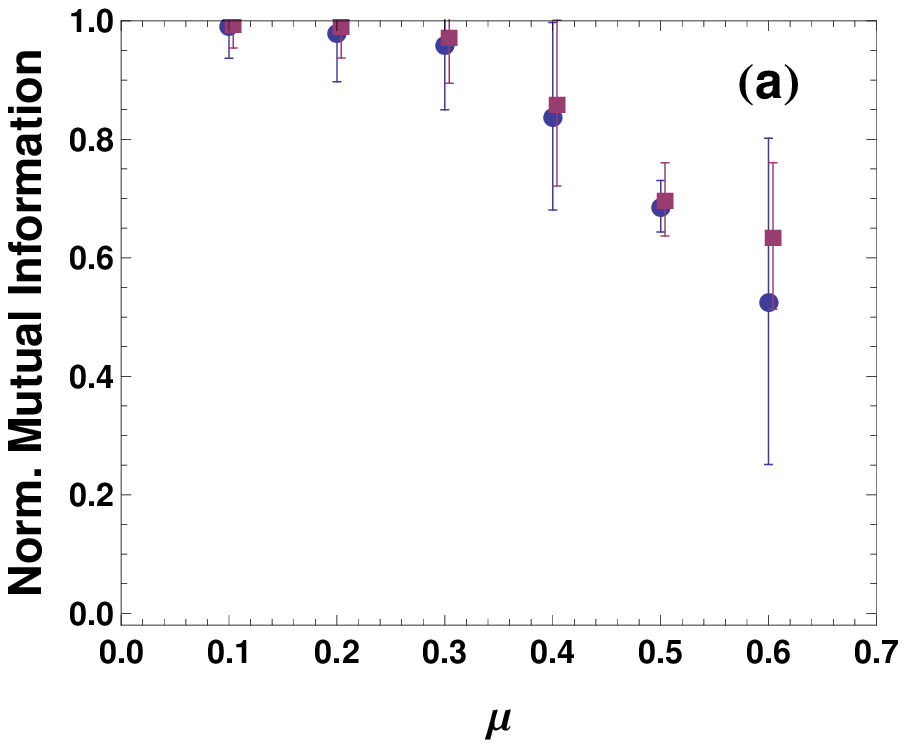}
		\hfill
		\includegraphics[width=0.47\textwidth]{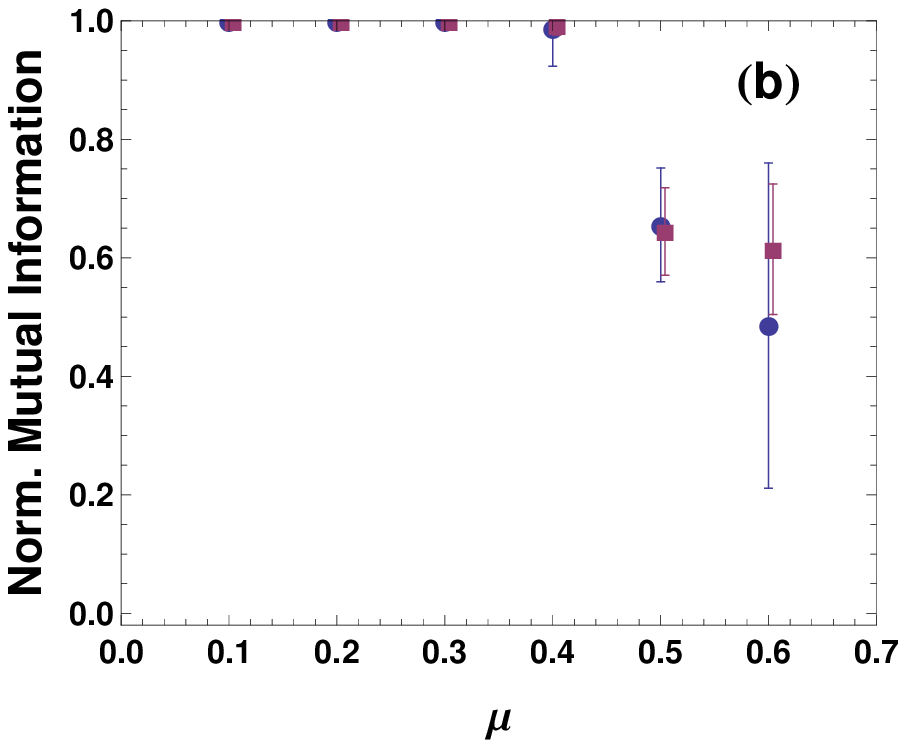}
	\caption{\label{fig:harel} Comparison of community detection efficiency between MERW (squares; $T=2$, $4$ iterations) and GRW (circles; $T=3$, $3$ iterations) transition matrix used in the first iteration of algorithm \cite{Harel} on benchmark graphs. Graph size:(a) $N=200$, (b) $N=1000$.}
\end{figure}

The algorithm consists in replacing edge weights of the original graph with the elements of the similarity matrix,
so that external (intercommunity) links get smaller weights, and the internal ones get larger weights.
The procedure is iterated until the differences between weights become large enough,
and the weights below a given threshold can be disposed of. What remains is the communities.
It is viable to use the transition matrix of MERW only in the first iteration step.
As illustrated in Fig.\ref{fig:harel}, MERW produces slightly better results, especially for considerable $\mu$. (Details of the comparisons are described in Sec.\ref{sec:bench}.)

\begin{figure}[hp]
	\centering
		\includegraphics[width=0.47\textwidth]{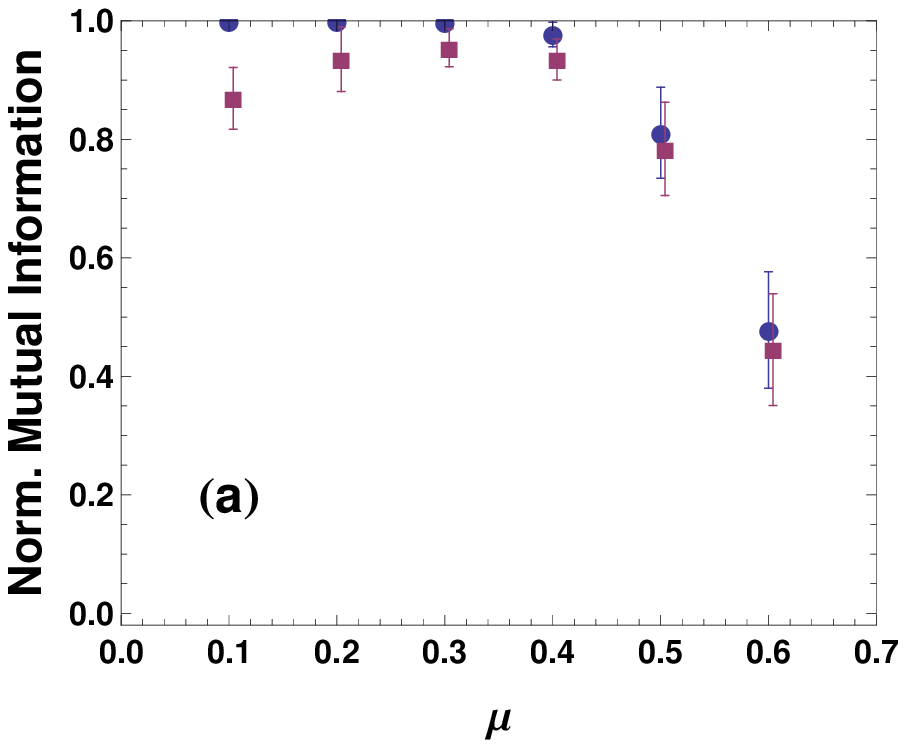}
		\hfill
		\includegraphics[width=0.47\textwidth]{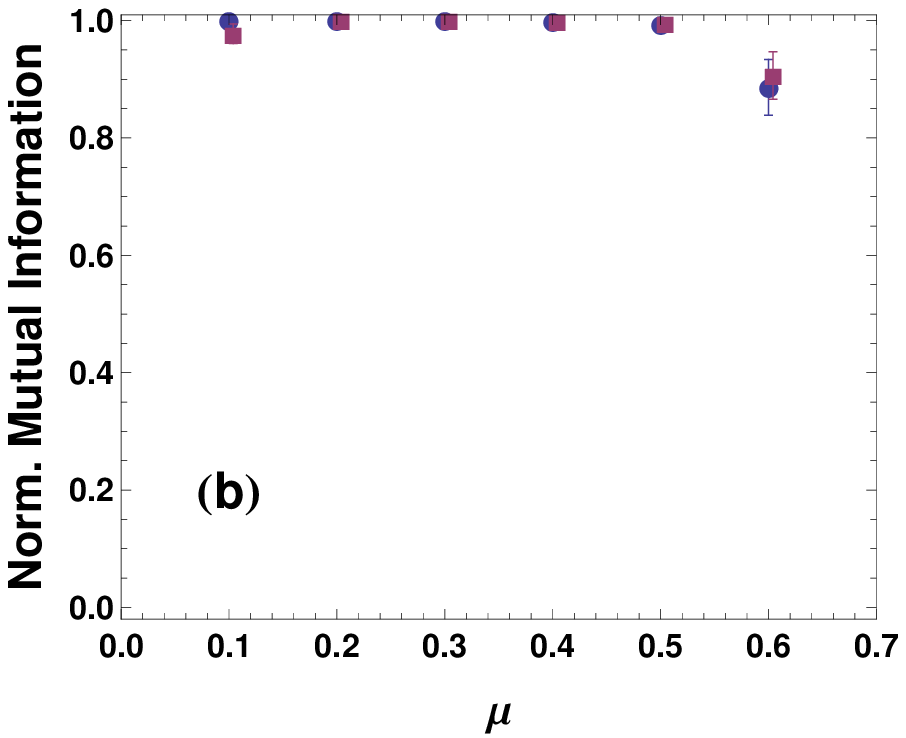}
	\caption{\label{fig:latapy} Comparison of community detection efficiency between MERW (squares; summed powers of $\mathbf{P}^t$, $t=1-3$) and GRW (circles; $t=3$) for the algorithm Pons and Latapy \cite{Latapy}. Graph size:(a) $N=200$, (b) $N=1000$.}
\end{figure}

Next, Pons and Latapy \cite{Latapy} introduced an algorithm using as a distance matrix between nodes of the graph the quantity given in \eqref{rdist}. Fig.\ref{fig:latapy}. MERW considerably decreases efficiency of the algorithm for small $\mu$; the precise reasons for that are not established.

In \eqref{Gmu}, the weights $e^{-\mu t}$ produce the resolvent operator of $\mathbf{A}$, but also factorial weights $\beta^t/t!$ might be introduced \cite{Estrada1,Estrada2}, yielding the heat kernel.
To analyse the resulting matrix one needs to remove the zeroth eigenmode of $\mathbf{A}$, so that $\mathbf{G}$ is well-defined.
The choice $e^{\mu}=\lambda_0$ is directly related to MERW.

\begin{figure}[hp]
	\centering
		\includegraphics[width=0.47\textwidth]{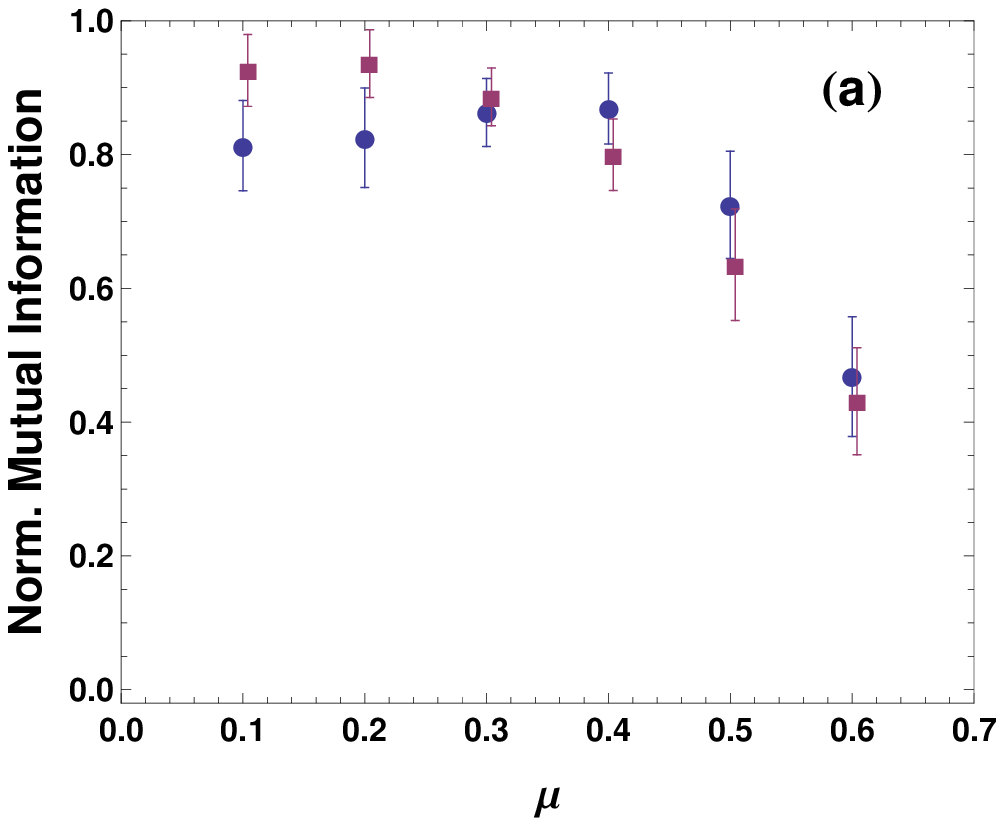}
		\hfill
		\includegraphics[width=0.47\textwidth]{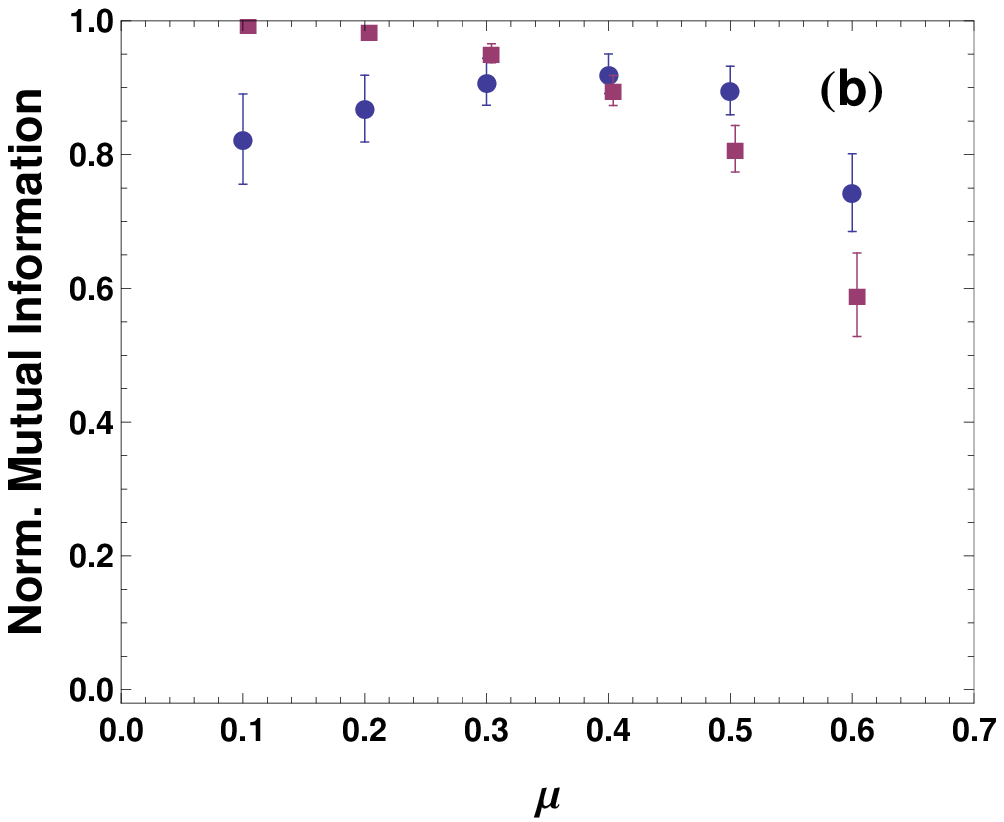}
	\caption{\label{fig:estrada} Comparison between $\lambda_0^t$ (MERW, squares) and $t!$ (circles) path weights. Graph size:(a) $N=200$, (b) $N=1000$.}
\end{figure}

The procedure \cite{Estrada1,Estrada2} goes on, producing a matrix with $0$s and $1$s in place of negative and positive entries of $\mathbf{G}$. The original idea involved finding all maximal cliques (maximal complete subgraphs) of the graph represented by this matrix. Since this is computationally strenuous, we use a much simpler approach and carry out hierarchical clustering on that matrix. To obtain communities, we take the dendrogram section which maximises modularity\cite{NG}. This algorithm, however, should be considered as only a very rough approach, just for the sake of preliminary comparison.
It can be seen in Fig.\ref{fig:estrada} that exponential weights works better for small $\mu$, while factorial weights give a reasonable performance for larger values of mixing parameter.

\begin{figure}[h]
	\centering
		\includegraphics[width=0.47\textwidth]{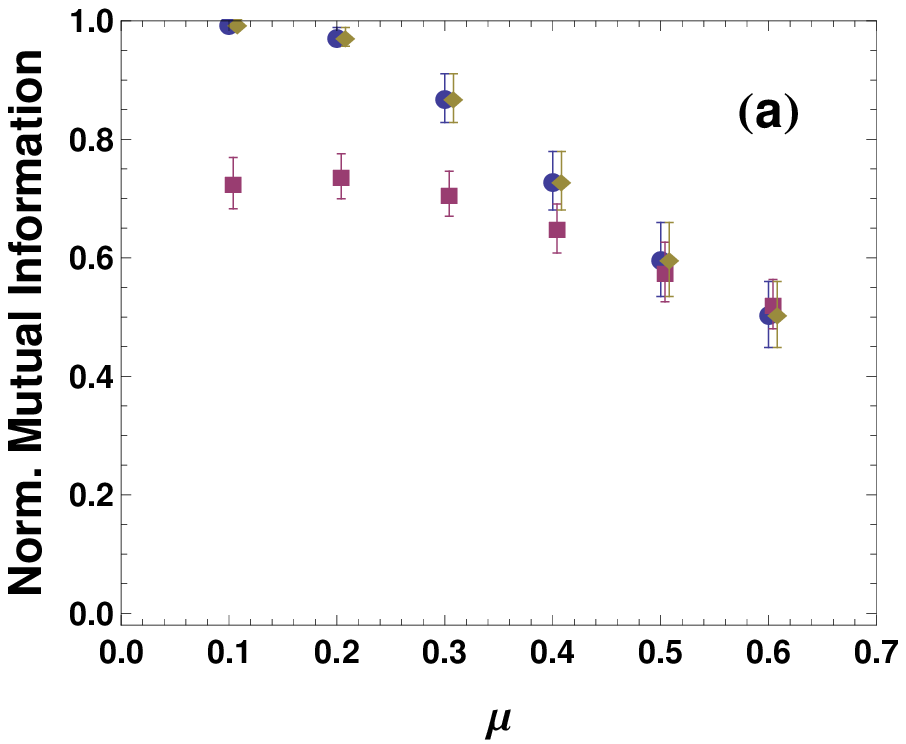}
		\hfill
		\includegraphics[width=0.47\textwidth]{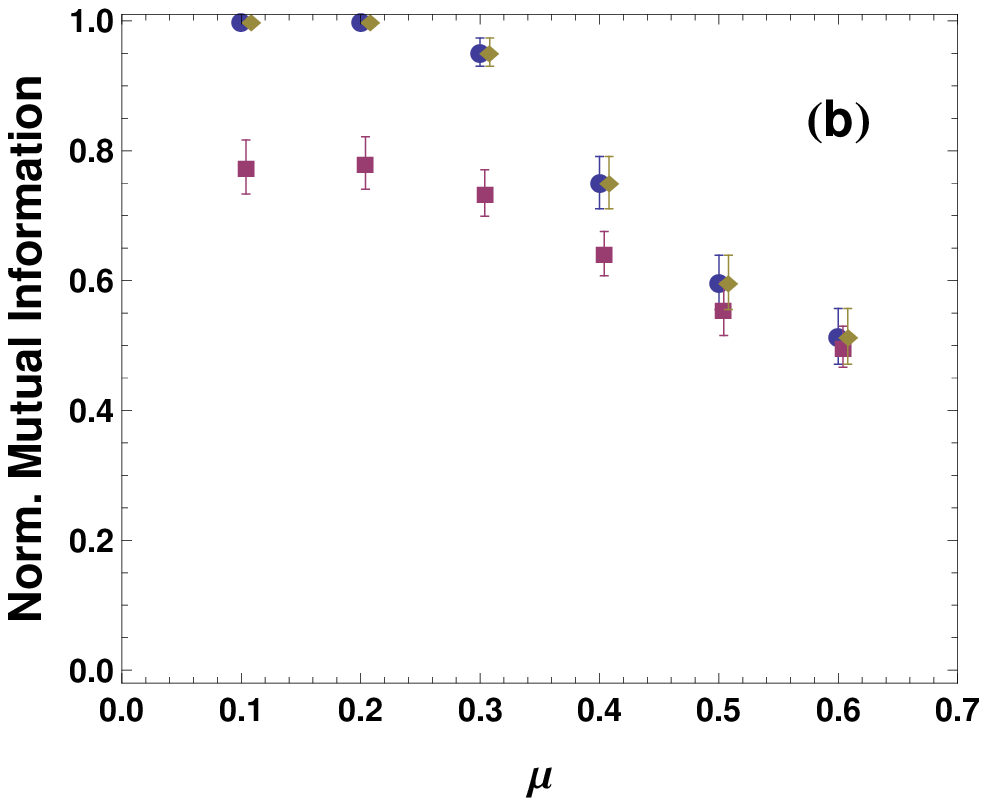}
	\caption{\label{fig:netwalk} Comparison between \emph{Netwalk} \cite{Zhou3} using MERW (squares), GRW (circles) and biased RW (diamonds)on benchmark graphs. Graph size:(a) $N=200$, (b) $N=1000$.}
\end{figure}

Lastly, one may look at methods grouping nodes according to their MFPT values. 
In \cite{Zhou1,Zhou3} a similarity matrix is introduced that computes the total of differences between MFPTs of random walkers incoming to given nodes $a$ and $b$ from any initial node
\beq
\Lambda_{ab}=\frac{\sqrt{\sum_{c\neq a,b}^{N} |M_{ac}-M_{bc}|^2}}{N-2}.
\eeq
On this basis the authors developed an algorithm called \emph{Netwalk}. We skip the details of the algorithm and refer the reader to the original papers.
In this case, the outcome of the comparison between MFPTs of different random walks (we also implement a biased random walk used originally by \emph{Netwalk}), in Fig.\ref{fig:netwalk}, shows that MERW should not be used in this algorithm. The original algorithm, however, works well only for very small $\mu$, and in general its performance is unexpectedly unreliable even for large network size.

\subsection{Benchmark graphs}
\label{sec:bench}

The algorithms in Sec.\ref{sec:community} are compared with the use of unweighted undirected LFR benchmark graphs introduced in \cite{Fortunato1} in a manner analogous to the authors' later work \cite{Fortunato2}. We take $100$ benchmark graphs with $N=200, 1000$ nodes; their exponents for the degree distribution and for the community size distribution are respectively $\tau_1=-2$ and $\tau_2=-1$. For $N=200$ the parameters are: the average degree of $10$, maximum degree of $30$, and the minimum and maximum community sizes are taken to be $5$ and $35$.
For $N=1000$: the average degree of $20$, maximum degree of $50$, and the minimum and maximum community sizes are $20$ and $100$, respectively.
The mixing parameter $\mu$ is the fraction of links a given node shares with the nodes outside its community. The parameter is approximately equal for all nodes in a graph, and its values are set to $\mu=0.1-0.6$. For the upper bound, most of the algorithms start to have severe problem with detecting communities. To check how good partition has been found we use the normalised mutual information \cite{Arenas} with respect to the partition planted in the benchmark. Let us note that the definition of a community here relies on the \emph{planted partition model}, which means that the performance of algorithms is checked in accordance with this particular definition.

\section{Conclusions}

We have briefly introduced the concept of maximal-entropy random walk and reviewed some of its features,
while in the main body of this paper we compared the performance of several community finding algorithm,
in which MERW-based (dis)similarity matrices substituted the original ones.

The results obtained by the most reliable method checked here, made by Latapy and Pons,
are comparable for GRW and MERW, although we note significant worsening for small networks when using the latter.

The other methods have not been previously compared on LFR benchmark graphs.
The one by Harel and Koren is generally not reliable for $\mu>0.4$. However, its performance is slightly improved by MERW for both small and large networks.
In contrast, MERW is not suited for \emph{Netwalk}.
Even for GRW, which was used originally,
this algorithm produces a markedly unsatisfactory results for the medium range of the mixing parameter in comparison to available state-of-the-art methods. 
The method based on factorial path weighting has considerable problems for small $\mu$. Surprisingly, switching to exponential weighting, which corresponds to MERW, produces better results than \emph{Netwalk}. In general, it performs reasonably well, even though the algorithm used simple hierarchical clustering as temporary means for the sake of comparison.

Whereas MERW exhibits surprising localisation and relaxation properties on some defective regular graphs,
this case study shows that on the LFR benchmark graphs, which are locally random,
this random walk can offer a performance of community finding methods comparable to that of GRW.
It remains to be investigated, if the behaviour of MERW on other types of graphs, including real-world networks,
is more distinctive. Further effort is also needed to determine whether development of a dedicated algorithm which makes better use of the information contained in this type of random walk is possible.

\section{Acknowledgement}
Project operated within the Foundation for Polish Science International Ph.D. Projects Programme co-financed by the European Regional Development Fund covering, under the agreement no. MPD/2009/6, the Jagiellonian University International Ph.D. Studies in Physics of Complex Systems.

\end{document}